\documentclass{emulateapj}
\usepackage{times}

\slugcomment{Submitted April 17, 2006; accepted August 4, 2006}

\newcommand{\eq}[1]{\begin{equation} #1 \end{equation}}

\newcommand\about {\hbox{$\sim$}}
\newcommand{\E}[1]{\hbox{$10^{#1}$}}
\newcommand\x     {\hbox{$\times$}}
\newcommand\mic    {\hbox{$\mu$m}}
\newcommand\Mo    {\hbox{$M_{\odot}$}}
\newcommand\Mc    {\hbox{$M_{\rm cl}$}}
\newcommand\mH    {\hbox{$m_{\rm p}$}}
\newcommand\NH    {\hbox{$N_{\rm H}$}}
\newcommand\nc    {\hbox{$n_{\rm cl}$}}
\newcommand\Nc    {\hbox{$N_{\rm cl}$}}

\newcommand\No    {\hbox{${\cal N}_0$}}
\newcommand\vc    {\hbox{$v_{\rm cl}$}}
\newcommand\Ri    {\hbox{$R_{\rm i}$}}
\newcommand\Mw    {\hbox{$\dot M_{\rm w}$}}
\newcommand\Mwc   {\hbox{$\dot M_{\rm w,cl}$}}
\newcommand\Macc  {\hbox{$\dot M_{\rm acc}$}}
\newcommand\MBH   {\hbox{$M_{\bullet\,7}$}}
\newcommand\Mtot  {\hbox{$M_{\rm tot}$}}
\newcommand\kms   {\hbox{km\,s$^{-1}$}}
\newcommand\rpc   {\hbox{$r_{\rm pc}$}}
\newcommand\cc    {\hbox{cm$^{-3}$}}
\newcommand\cs    {\hbox{cm$^{-2}$}}
\newcommand\Myr   {\hbox{$\Mo\,{\rm yr}^{-1}$}}
\newcommand\erg   {\hbox{erg\,s$^{-1}$}}

 \shorttitle{The AGN Torus}
 \shortauthors{Elitzur \& Shlosman}

\begin{document}

\title{The AGN Obscuring Torus --- End of  the ``Doughnut'' Paradigm?}

\author{Moshe Elitzur\altaffilmark{1} and Isaac Shlosman}
\altaffiltext{1}{On sabbatical leave at Laboratoire d'AstrOphysique de
Grenoble}

\affil{Department of Physics and Astronomy, University of Kentucky,
                   Lexington, KY 40506-0055
}
\email{moshe@pa.uky.edu, shlosman@pa.uky.edu}

\begin{abstract}
Unified schemes of active galactic nuclei (AGN) require an obscuring dusty
torus around the central engine. The compact sizes (only a few pc) determined
in recent high-resolution observations require that the obscuring matter be
clumpy and located inside the region where the black-hole gravity dominates
over the galactic bulge. This location is in line with the scenario depicting
the torus as the region of the clumpy wind coming off the accretion disk in
which the clouds are dusty and optically thick. We study here the outflow
scenario within the framework of hydromagnetic disk winds, incorporating the
cloud properties determined from detailed modeling of the IR emission from
clumpy tori. We find that torus clouds were likely detected in recent water
maser observations of NGC 3079. In the wind scenario, the AGN main dynamic
channel for release of accreted mass seems to be switching at low luminosities
from torus outflow to radio jets. The torus disappears when the bolometric
luminosity decreases below \about\ \E{42} \erg\ because the accretion onto the
central black hole can no longer sustain the required cloud outflow rate. This
disappearance seems to have been observed in both LINERs and radio galaxies.
With further luminosity decrease, suppression of cloud outflow spreads radially
inward from the disk's dusty, molecular region into its atomic, ionized zone,
resulting in disappearance of the broad emission line region at lower
luminosities, yet to be determined.

\end{abstract}


\section{Introduction}

The great diversity of AGN classes has been explained by a single unified
scheme (e.g., Antonucci 1993; Urry \& Padovani 1995). The nuclear activity is
powered by a super\-massive (\about\E6--\E{10} \Mo) black hole (SBH) and its
accretion disk. This central engine is surrounded by dusty clouds, which are
individually optically thick, in a toroidal structure (Krolik \& Begelman
1988). The clumpy torus is characterized by two fundamental properties: (1)
Anisotropic obscuration of the central region so that sources viewed face-on
are recognized as type 1 objects, those observed edge-on are type 2. From the
statistics of Seyfert 1 and 2 galaxies, the torus height\footnote{The highest
point that gives $\tau_{\rm V} \sim 1$ along the line of sight.} and radius
obey $H/R$ \about\ 1 (Schmitt et al.\ 2001). (2) Dust re-emission in the IR of
the AGN obscured radiation. This emission is the only means to find the torus
radial extent since obscuration cannot determine separately $H$ and $R$.

Two approaches have been taken for the torus dynamic origin. A hydrostatic
scenario depicts the torus as a doughnut-like structure populated by molecular
clouds accreted from the galaxy (Krolik \& Begelman 1988). However, the origin
of vertical motions capable of sustaining the clouds in a hydrostatic structure
with $H\sim R$ was recognized from the start as problematic and has eluded
solution thus far (e.g., Davies et al 2006). The other scenario, based on the
seminal work by Blandford \& Payne (1982), involves the outflow of clouds
embedded in a hydromagnetic disk wind (Emmering et al.\ 1992, EBS hereafter;
K\"onigl \& Kartje 1994; Kartje \& K\"onigl 1996; Bottorff et al.\ 1997, 2000;
Kartje at al.\ 1999; Everett 2005). In this approach the torus is merely a
region in the wind which happens to provide the required toroidal obscuration,
i.e., it is that region wherein the clouds are dusty and optically thick.

Recent high-resolution IR observations indicate that the torus size might be no
more than a few pc (Elitzur 2005, and references therein); in particular, VLTI
observations of NGC 1068 show that the FWHM size of the 12 \mic\ emission is
only \about\ 4 pc (Jaffe et al 2004). The compact sizes place the torus inside
the region where the SBH gravity dominates over the galactic bulge. Consider a
bulge that induces a linearly rising rotation curve with $\Omega \sim
1$\,\kms\,pc$^{-1}$, as is typical of AGN host galaxies (Sofue et al.\ 1999),
and an SBH with a mass of \MBH\x\E7\,\Mo\ at its center. The SBH will dominate
the gravitational motions within a radius 35\,pc\x$(\MBH/\Omega_1^2)^{1/3}$,
where $\Omega_1 = \Omega/(1$\,\kms\,pc$^{-1})$. Since the torus is well within
this region, its dynamic origin is determined in all likelihood by the central
engine and its accretion disk, giving strong impetus to the outflow paradigm.
The central accretion disk appears to be fed by a midplane influx of cold,
clumpy material from the main body of the galaxy (Shlosman et al 1990, and
references therein). Approaching the center, conditions for developing
hydromagnetically- or radiatively-driven winds above this equatorial inflow
become more favorable. The disk-wind geometry provides a natural channel for
angular momentum outflow from the disk and is found on many spatial scales,
from protostars to AGN (Blandford \& Payne 1982; EBS; Ferreira 2006). Although
theoretical questions involving cloud-uplift and wind-driving mechanisms remain
(e.g., EBS; Ferreira 2006), observations give ample evidence for winds and
cloud motions (e.g., Arav et al 1997; Elvis 2004) and support for a disk-wind
geometry (e.g., Hall et al.\ 2003) in AGN.

Clumpiness provides a natural explanation for the compactness of the IR
emission from AGN tori, and detailed modeling of this emission sets
phenomenological constraints on the torus cloud properties (Nenkova et al.\
2002, 2006; Elitzur et al.\ 2004; Elitzur 2005). Here we explore whether the
torus outflow scenario can properly accommodate these constraints, and the
corollaries for the AGN paradigm.

\section{Cloud Properties}

The only property of individual clouds constrained from the IR modeling is
their optical depth. It should lie in the range \about\ 20--150 at the $V$
band, i.e., column densities \NH\ \about\ \E{22}--\E{23} cm$^{-2}$ assuming
standard dust-to-gas ratio. Clouds uplifted into the wind expand while moving
away from the disk. A cloud starting with \NH\ \about\ \E{23} cm$^{-2}$ ceases
to partake in obscuration when its column is reduced by \about\ 100. A cloud
starting with a smaller column will rise to a smaller height before leaving the
obscuring torus. Consider the tidal torque of the central SBH at a distance
\rpc\ in pc, where the Keplerian period is $t_{\rm K} = 3\x\E4 M_{\bullet
7}^{-1/2} r_{\rm pc}^{3/2}~{\rm yr}$. To prevent a cloud with density $\rho =
\mH n$ from shearing, it must be at least partially confined by its own gravity
and/or the ambient magnetic field $B$. In the former case, the characteristic
Jeans timescale is $t_{\rm J} \sim (G\rho)^{-1/2} = 3\x\E4 n_7^{-1/2}~{\rm
yr}$, where $n_7 = n/(\E7\,\cc)$. Therefore, resistance to tidal shearing
($t_{\rm J} < t_{\rm K}$) requires $n_7 > \MBH/r_{\rm pc}^3$, which, for a
cloud with column \NH\ = $N_{\rm H,23}\x\E{23}\,\cs$, implies a size $a \la
\E{16}N_{\rm H,23}r_{\rm pc}^3/\MBH$ cm. The corresponding cloud mass is $\Mc
\la 7\x\E{-3}N_{\rm H,23}a_{16}^2$\,\Mo, where $a_{16} = a/(\E{16}\,\rm cm)$.
The self-gravity of such a low mass cannot hold it together against dispersal
by any reasonable internal velocities. However, an external magnetic field $B
\sim 1.5\,\sigma_5n_7^{1/2}$ mG would suffice if the internal velocity
dispersion is $\sigma_5\x$1\,\kms.

Clouds with these very same properties may have been detected by Kondratko et
al.\ (2005) in recent H$_2$O maser observations of NGC~3079. Individual maser
features detected in this source have sizes $\la$ 2\x\E{16} cm and span the
radial range \about\ 0.4--1.3 pc. Similar to other objects, most maser features
in this source reside in an edge-on rotating disk. However, for the first time,
four features were found significantly out of the disk plane yet their
line-of-sight velocities reflect the velocity of the most proximate side of the
disk. Kondratko et al.\ note that this can be explained if, as proposed by
Kartje et al.\ (1999), maser clouds rise to high latitudes above the rotating
structure while still maintaining, to some degree, the rotational velocity
imprinted by the parent disk. Because the detected maser emission involves
cloud-cloud amplification that requires precise alignment in both position and
velocity along the line-of-sight, the discovery of four high-latitude maser
features implies the existence of many more such clouds partaking in the
outflow. {\em We suggest that the high-latitude masers are yet another
manifestation of the dusty, molecular clouds that make up the torus region of
the disk-wind.} Densities of H$_2$O masers are frequently quoted as \about\
\E8--\E9\ \cc, but these are the optimal densities to produce the highest
possible brightness from a single cloud. Because of the scaling properties of
H$_2$O pumping (Elitzur et al.\ 1989), maser clouds with $n$ \about\ \E7\ \cc\
and \NH\ \about\ \E{23} \cs\ produce near-optimal inversion and detected
radiation in cloud-cloud amplification.

To produce a torus with the required thickness, the column density of a cloud
must remain above \about\ \E{21} \cs\ during the rise to height $H \sim r$. If
the cloud expands at the speed of sound $c_{\rm s}$ while rising with velocity
$v$, then its size will increase by a factor $x \simeq 1 + (r/a)(c_{\rm s}/v)$.
We expect $c_{\rm s}/v \simeq \E{-2}$ since $c_{\rm s}$ is 1\,\kms\ at a
temperature of 100 K and the velocity scale of the cloud outflow is expected to
be comparable to the Keplerian velocity, $208(\MBH/\rpc)^{1/2}$\,\kms.  With
$a/r \sim \E{-3}$, the 3-D expansion of such a cloud will cause its size to
increase by a factor $x$ \about\ 10 and the column density ($\propto \Mc/a^2$)
to decrease within the torus region by a factor $x^2$ \about\ 100, as required.
Furthermore, in a cloud-loaded hydromagnetic disk wind the ambient magnetic
field suppresses cloud expansion across the field lines while guiding the cloud
motion. Magnetic confinement can be efficient in either 3-D, e.g., the ``melon
seeds'' model of diamagnetic clouds (Rees 1987), or 2-D, e.g., the slender
tubes model of paramagnetic clouds (EBS). In the simplest picture of constant
area flux tubes and thermal expansion only along the field lines, the cloud
column density remains unchanged in the longitudinal direction while decreasing
as $1/x$ in the orthogonal direction. In reality, the decrease of column
density is likely to fall somewhere between the $1/x^2$ of 3-D thermal
expansion and $1/x$ of constant field. Whichever the case, magnetic effects
always supplement the thermal confinement, enabling smaller and denser torus
clouds.

\section{Mass Outflow}

At a bolometric luminosity $L = L_{45}$\x\E{45}\,\erg, the inner radius of the
torus cloud distribution is \Ri\ $\simeq$
0.4\,pc\x$L_{45}^{1/2}T_{1500}^{-2.6}$, where $T_{1500}$ is the dust
sublimation temperature normalized to 1500 K. This distance was determined from
the dust temperature on the illuminated face of an optically thick cloud of
composite dust representing standard mix of Galactic grains (Nenkova et al
2006). Some dust components can exist closer than \Ri, and attenuation of the
central UV radiation could also reduce \Ri\ at low latitudes. The torus outer
radius is $Y\Ri$, with $Y$ \about\ 5--10. The quantity constrained by IR
modeling is the number of clouds per unit length, $\Nc = \nc A_{\rm cl}$, where
\nc\ is the number of clouds per unit volume and $A_{\rm cl}$ the cloud
cross-sectional area. Neither \nc\ nor $A_{\rm cl}$ enter independently into
the calculation of the torus emission. An analytic expression for
$\Nc(r,\beta)$ in terms of radial distance $r$ and angle $\beta$ from the
equatorial plane that agrees reasonably with all observations, and can serve as
a ``standard'' cloud distribution, is $\Nc \propto \No\, r^{-2}
\exp(-\beta^2/\sigma^2)$. Here \No\ (\about\ 5--10) is the total number of
clouds encountered, on average, along a radial equatorial ray, and $\sigma$
(\about\ 45\degr) measures the angular width of the cloud distribution. This
phenomenological parametrization of the torus cloud distribution was not meant
to account for the underlying disk, where the clouds originate, and should hold
only at inclinations $\beta$ above the disk surface. In practice, we neglect
the disk thickness and utilize this expression at all $\beta$.

The mass density of clouds is \Mc\nc = \mH\NH\Nc, and the cloud outflow rate
generating the torus is $\mH\NH\int \Nc\,\vc\,dA$. The integration is over the
disk area where the torus clouds are injected out of the plane with velocity
\vc. This velocity can be estimated from the dispersion velocity of molecular
material in the disk --- an outflow is established when the velocity of the
ordered motion becomes comparable to that of the local random motions. From the
maser observations in NGC~3079 Kondratko et al.\ (2005)
find that the velocity dispersion in a small region ($\le$ 5\x\E{16} cm) of
strong emission is only \about\ 14 \kms, which can be considered typical of the
local random motions. Therefore, we parameterize the injection velocity as
$\vc(r) = 10\,\kms\x v_6 u(r)$ where $u = v(r)/v(\Ri)$ and $v_6$ is the
velocity at the inner radius \Ri\ normalized to 10 \kms; both $v_6$ and $u(r)$
are expected to be of order unity. The torus mass outflow rate in the form of
clouds is thus
\eq{\label{eq:Mdot1}
  \Mwc = 0.02\,L_{45}^{1/2}\,T_{1500}^{-2.6}\,
            N^{\rm tot}_{H,23}\,v_6\,\x\,I_1\ \Myr.
}
Here $N^{\rm tot}_{H,23} = \No N_{H,23}$ is the column density through all
clouds in the equatorial plane in units of \E{23}\,\cs\ and $I_1 = \int_1^Y
(u/y) dy\x Y/(Y - 1)$ is an unknown factor of order unity. The mass inflow rate
across the torus is related to the bolometric luminosity via $\Macc =
0.02\,L_{45}/\epsilon$ \Myr, where $\epsilon$ is the accretion efficiency at
the torus radii. Note that \Macc\ is a function of $r$, since the mass loss in
a hydromagnetic wind can be a significant fraction of the disk accretion rate;
at radii characteristic of the torus, \Macc\ can be substantially larger, and
$\epsilon$ substantially smaller, than in the innermost accretion disk (EBS).
The results for \Mwc\ and \Macc\ yield
\eq{\label{eq:Mdot2}
   {\Mwc\over\Macc} = \epsilon N^{\rm tot}_{H,23} v_6 L_{45}^{-1/2}
         T_{1500}^{-2.6} \x\,I_1.
}
When $L$ decreases, \Mwc/\Macc\ increases. However, this ratio cannot exceed
unity. Moreover, the disk wind applies a torque on the underlying accretion
disk because each magnetic field line can be considered rigid from its
footpoint $r_0$ and up to the Alfv\'en radius, $r_A$. For a {\em total} outflow
rate \Mw\ of a continuous wind loaded with clouds, angular momentum
conservation gives $\gamma \equiv \Mw/\Macc \sim (r_0/r_A)^2$, with $\gamma$
\about\ 0.1--0.25 either constant (EBS) or weakly dependent on $r_0$ (Pelletier
\& Pudritz 1992). Therefore, $\Mwc \la \gamma\Macc$. When the luminosity is
decreasing, eventually this bound is violated by eq.\ \ref{eq:Mdot2}\ and the
system can no longer sustain the cloud outflow rate required by the torus wind.
That is, {\em the torus should disappear in low-luminosity AGN}. If we take
$\gamma \simeq 0.2$, an accretion efficiency $\epsilon \simeq 0.01$ (EBS) and
all other parameters \about\ 1, the torus disappearance should occur at
bolometric luminosities below \about\ \E{42} \erg. The timescale for this is
short. The overall mass in torus clouds is $\Mtot = \mH\NH\int \Nc\,dV \simeq
\E{3}\,N^{\rm tot}_{H,23} L_{45} Y \Mo$, so the depletion time scale when \Mwc\
$\simeq \gamma$\Macc\ is $\Mtot/\Mwc \simeq 5\x\E4\,(\epsilon/\gamma) Y N^{\rm
tot}_{H,23}$ years. Since $\epsilon Y/\gamma$ \about\ 1, the torus will
disappear within a few Keplerian orbits.

The kinetic luminosity $\dot E_{\rm k}$ of the torus cloud outflow is generally
insignificant. A calculation similar to that for \Mwc\ yields $\dot E_{\rm k} =
7\x\E{35} L_{45}^{1/2}N^{\rm tot}_{H,23} v_6^3\x\,I_3\ \erg$, where $I_3 =
\int_1^Y (u^3/y) dy\x Y/(Y - 1)$ is of order unity. This is negligible in the
overall energy budget of the AGN. Finally, the volume averaged density of the
clouds is $\phi n$, where $\phi = a\Nc$ is their volume filling factor. The
initial outflow velocity of clouds is roughly 10 times larger than their
internal velocity dispersion. Since we have $\phi$ \about\ \E{-2}, the average
kinetic energy is similar for the cloud outflow and the internal motions.
Therefore the assumption that the field limiting the clouds expansion is also
guiding their motion is self-consistent.

\section{Summary and Discussion}

The analysis presented here shows that the disk wind scenario is compatible
with cloud properties inferred from models of the torus IR emission.
Furthermore, the very same dusty, molecular clouds uplifted from the disk
appear to have been detected in water maser observations of NGC 3079 (Kondratko
et al.\ 2005). Proper-motion measurements, if possible, of these high-latitude
masers would be extremely valuable. Detection of maser polarization seems less
likely because it would require proper alignment of the magnetic fields among
maser clouds that amplify each other (Elitzur 1996). Co-location of water
masers and the torus seems to be observed also in NGC 1068, where, as noted by
Poncelet et al.\ (2006), the two are at similar radial distances. The masers
there trace a thin, edge-on rotating disk between 0.65 and 1.1 pc (Greenhill \&
Gwinn 1977; Gallimore et al.\ 2001). It is intriguing that molecular structures
somewhat resembling the maser disk, and which might be its extension, are seen
also on much larger scales in NGC 1068. Galliano et al.\ (2003) find that H$_2$
and CO emissions come from clouds in a disk close to edge-on with radius 140 pc
and scale height 20 pc, for $H/R$ \about\ 0.15. From CO mapping, Schinnerer et
al.\ (2000) conclude that $H$ \about\ 9--10 pc at radial distance of 70 pc, for
a similar $H/R$. Detailed, high-resolution mapping of the entire region between
\about\ 1 and 50 pc around the nucleus can shed light on the variation in the
disk structure during the transition from gravitational domination by the
galactic bulge to the black hole, and on the origin of the disk wind.

Detectable only in edge-on AGN, the water masers reside in the inner molecular
regions of the accretion disk. At smaller radii, the disk composition switches
from dusty and molecular to atomic and ionized (e.g., EBS; Kartje et al 1999).
From time variations of X-ray obscuration that indicate cloud motions across
the line-of-sight, Risaliti et al.\ (2002, 2005) find that some of the
obscuring clouds are at radii considered typical of the AGN broad-line region
(BLR). Clouds at the very inner edge of the molecular disk are protected there
from the central UV. Upon injection into the wind and exposure to the AGN
radiation, these clouds can be responsible for the broad emission lines at low
altitudes above the disk, and for the warm absorption at higher altitudes.
Indeed, the kinematics of such clouds has been shown to fit nicely the BLR line
profiles (EBS). Clouds uplifted further out will have their entire trajectories
at $r >$ \Ri\ and form the obscuring torus. Since cloud properties must span
some range, the torus structure is likely stratified. Clouds that reach the
maximal height $H \sim r$ must start with a size of \about\ \E{16} cm and
column density of \about\ \E{23} \cs. Clouds with smaller initial size or
column will lose the required obscuration at lower heights.

The fraction of type~2 objects appears to decrease with $L$, in accordance with
the ``receding torus'' model (Simpson 2005).  A decrease of $H/R$ with $L$
arises naturally in the torus outflow scenario. The torus radial dimensions are
set from its inner radius \Ri\ and increase with $L$ due to dust sublimation.
On the other hand, since outflow velocities are expected to decrease with $r$,
thermal expansion will reduce the cloud column density below the range of torus
clouds at $H < r$ if the relative size $a/r$ is roughly constant. An additional
possible factor limiting the height at larger $L$ is bending of the streamlines
by radiation pressure (e.g., de Kool \& Begelman 1995).

A key prediction of the wind scenario is that the torus disappears at low
bolometric luminosities ($\la$ \E{42} \erg) because mass accretion can no
longer sustain the required cloud outflow rate, i.e., the large column
densities. Observations seem to corroborate this prediction. In an HST study of
a complete sample of low-luminosity ($\la$ \E{42} \erg) FR I radio galaxies,
Chiaberge et al.\ (1999) detected the compact core in 85\% of sources and
argued that this high detection rate implies the absence of an obscuring torus.
This finding was enhanced when Whysong \& Antonucci (2004 and references
therein) demonstrated that M87, one of the sample sources, indeed does not
contain an obscuring torus by placing stringent limits on its thermal IR
emission; a similar conclusion was reached by Perlman et al.\ (2001).
Additional evidence comes from UV monitoring of LINERs with $L \la$ \E{42}
\erg\ by Maoz et al.\ (2005). By detecting variability in most of the monitored
objects they demonstrated that the AGN makes a significant contribution to the
UV radiation in each source and that it is relatively unobscured in both type 1
and type 2 LINERs. The histogram of UV colors shows an overlap between the two
populations, with the difference between the peaks corresponding to dust
obscuration in the type 2 LINERs of only \about\ 1 magnitude in the R band.
Such toroidal obscuration is minute in comparison with higher luminosity AGN.

The predicted torus disappearance at low $L$ does not imply that the cloud
component of the disk wind is abruptly extinguished, only that its outflow rate
is less than required by the ``standard'' IR emission observed in quasars and
high-luminosity Seyferts. When \Mwc\ drops below ``standard'' torus values,
which at each AGN might occur at somewhat different $L$ because of the
intrinsic spread of parameters in eq.\ \ref{eq:Mdot2} ($\epsilon$, $v$ etc),
the outflow still provides toroidal obscuration as long as its column exceeds
\about\ \E{21} \cs. Indeed, Maoz et al find that some Liners do have
obscuration, but much smaller than ``standard''. Line transmission through a
low-obscuration torus might also explain the low polarizations of broad
H$\alpha$ lines observed by Barth et al (1999) in some low luminosity systems.
The properties of thermal IR emission from such low luminosity AGN should
differ from the ``standard'', a prediction that should be tested in
observations.

If the toroidal obscuration were the only component removed from the system,
all low luminosity AGN would become type 1 sources. In fact, among the LINERs
monitored and found to be variable by Maoz et al (2005) there were both sources
with broad H$\alpha$ wings (type 1) and those without (type 2). Since all
objects are relatively unobscured, the broad line component is truly missing in
the type 2 sources in this sample. Similarly, Laor (2003) presents arguments
that some ``true'' type 2 sources, i.e., having no obscured BLR, do exist among
AGNs with $L \la \E{42}\,\erg$ (see also Ho et al 1997). Both findings have a
simple explanation if when $L$ decreases further, the suppression of mass
outflow spreads radially inward from the disk's dusty, molecular region into
its atomic, ionized zone. Then the torus disappearance, i.e., removal of the
toroidal obscuration by the dusty wind, would be followed by a diminished
outflow from the inner ionized zone and disappearance of the BLR at lower,
still to be determined, luminosities.

Within the framework of a hydromagnetic wind, $\Mw \propto \Macc$, and
expressing the luminosity in terms of \Macc\ in eq.\ \ref{eq:Mdot1}, $\Mwc
\propto (\epsilon\Macc)^{1/2}$. The torus cloud and total outflow rates both
vary in the same direction as the mass accretion rate. In contrast, Ho (2002)
finds that the radio loudness of AGN is {\em inversely} correlated with the
mass accretion rate (see also Greene et al.\ 2006). That is, when \Macc\ is
decreasing, the torus cloud and total outflow rates are decreasing too while
the radio loudness is increasing. It seems that the AGN switches its main
dynamic channel for release of accreted mass from torus outflow at higher
luminosities to radio jets at lower ones, with a certain degree of overlap. As
noted also by Greene et al, X-ray binaries display a similar behavior,
switching between radio quiet states of high X-ray emission and radio loud
states with low X-ray emission (see Fender et al.\ 2004).

\acknowledgments We have greatly benefited from discussions with many
colleagues, but wish to single out for special thanks remarks by Reinhard
Genzel, Chris Henkel, Ari Laor, Dani Maoz. This work was performed while M.E.\
spent a most enjoyable sabbatical at LAOG, Grenoble. Partial support by NASA
and NSF is gratefully acknowledged.

\end{document}